\documentclass{aa}
\input epsfig.sty

\def\ltsima{$\; \buildrel < \over \sim \;$}
\def\simlt{\lower.5ex\hbox{\ltsima}}
\def\gtsima{$\; \buildrel > \over \sim \;$}
\def\simgt{\lower.5ex\hbox{\gtsima}}

\begin{document}
   \thesaurus{(03) 13.25.2 -- 11.09.1 -- 11.19.1}
  \title{BeppoSAX observations of Narrow-line Seyfert 1 galaxies: I. Ton~S~180}

   \author{A. Comastri 
              \inst{1}
     \and      F. Fiore
              \inst{2,3}
     \and      M. Guainazzi
              \inst{3}
     \and      G. Matt 
              \inst{4}
     \and      G.M. Stirpe
              \inst{1}
     \and      G. Zamorani
              \inst{1}
     \and      W.N. Brandt
              \inst{5}
     \and      K.M. Leighly
              \inst{6}
     \and      L. Piro
              \inst{7}
     \and      S. Molendi
              \inst{8}
     \and      A.N. Parmar
              \inst{9}
     \and      A. Siemiginowska
              \inst{10}
     \and      E.M. Puchnarewicz
              \inst{11}
}
   \offprints{A. Comastri (comastri@astbo3.bo.astro.it)}

  \institute {Osservatorio Astronomico di Bologna, Via Zamboni 33, I--40126
               Bologna, Italy
    \and  Osservatorio Astronomico di Roma, Via dell'Osservatorio, I--00040
           Monteporzio--Catone, Italy
    \and  SAX--Science Data Center, Nuova Telespazio, Via Corcolle 19,
           I--00131 Roma, Italy
    \and  Dipartimento di Fisica ``E. Amaldi", Universit\`a degli Studi 
          ``Roma Tre", Via della Vasca Navale 84, I--00146 Roma, Italy
    \and  Dept. of Astronomy and Astroph., The Pennsylvania 
          State University, 525 Davey Lab, University Park, PA 16802, USA
    \and  Department of Astronomy, Columbia University, 538 West 120th Street,
          New York, NY 10027, USA   
    \and  Istituto di Astrofisica Spaziale -- C.N.R., via E. Fermi 21, 
           I--00044 Frascati, Italy
    \and  Istituto di Fisica Cosmica e Tecnologie Relative -- C.N.R.,
           via Bassini 15, I--20133 Milano, Italy
    \and  Astrophysics Division, Space Science Dept. of ESA ESTEC/SA, 
           NL--2200 AG Noordwijk, The Netherlands
    \and Harvard--Smithsonian Center for Astrophysics, 60 Garden St.,
            Cambridge, MA 02138, USA
    \and  Mullard Space Science Laboratory, University College London,
           Holmbury St. Mary, Dorking, Surrey RH5 6NT, UK             
}

   \date{Received / Accepted }

\titlerunning{BeppoSAX observations of NLS1: Ton S 180}
\authorrunning{A. Comastri et al.}

   \maketitle

   \begin{abstract}
We report on the first spectrum up to 10 keV of the bright 
narrow-line Seyfert~1 galaxy Ton~S~180, obtained with the imaging instruments
onboard BeppoSAX. This is the first observed source in a sample
of a dozen narrow-line Seyfert~1 galaxies 
in the BeppoSAX Core Program.
We also present and discuss 
a high quality optical spectrum taken at the 1.5~m telescope at 
La Silla two months before the BeppoSAX observation. 
 
The X--ray spectrum shows a clear hardening 
above about 2 keV, where a power law with 
$\Gamma \simeq 2.3$ plus an iron line
provide a good description of the data. 
This slope is significantly steeper than the typical one 
for classical Seyfert 1's and quasars.
The best fit line energy is suggestive of highly ionized iron,
which would support the idea that the high accretion rate is (one of)
the fundamental parameter(s) characterizing the Narrow Line Seyfert 1 
phenomenon.

 \end{abstract}

%

\section{Introduction}

Narrow-line Seyfert 1 galaxies (hereafter NLS1; see Osterbrock \& Pogge 1985)
are defined by their optical emission line properties. They lie at the
lower end of the broad line width distribution for the Seyfert~1 class
with typical values of the H$\beta$ FWHM in the range 500--2000 
km $s^{-1}$. The 
[O~{\sc iii}]/H$\beta$ ratio is $<$ 3, and 
strong Fe~{\sc ii}, Ca~{\sc ii} triplet $\lambda8498$, $\lambda8542$, 
$\lambda8662$ and O~{\sc i} $\lambda8446$
emission lines as well as high ionization lines typical of Seyfert~1 galaxies 
are common among these objects. 
A remarkable and extremely strong anti--correlation was found between 
ROSAT PSPC X--ray spectral slope (0.1--2.0 keV) and the H$\beta$ 
FWHM (Boller et~al. 1996, hereafter BBF96; Forster \& Halpern 1996; 
Grupe 1996; Laor et~al. 1997).
Steep soft X--ray spectral slopes with typical photon indices 
(derived from single power-law fits) in the range $\Gamma \simeq$ 3--5
are only found among objects with narrow optical permitted lines. 

Large-amplitude and rapid X-ray variability is 
common among NLS1. Moreover, there is 
evidence for giant-amplitude X-ray variability 
(up to about two orders of magnitude) in 
IRAS~13224--3809 (Boller et al. 1993, 1997; Otani et al. 1996),
PHL 1092 (Forster \& Halpern 1996),  
RE~J~1237+264 (Brandt et~al. 1995; Grupe et~al. 1995a) and
WPVS007 (Grupe et~al. 1995b). It is notable
that such extreme variability properties have been discovered, 
so far, only in NLS1.

In the harder $\approx$ 2--10~keV 
energy band, recent ASCA observations 
have shown that NLS1 can have very different behaviour from
classical broad-lined Seyfert~1s and quasars.
A comparative ASCA study of a 
large sample of NLS1 and broad-line Seyfert~1s 
revealed that the $\approx$~2--10~keV 
ASCA spectral slopes of NLS1 are
generally steeper than those of broad line objects at 
a high statistical confidence level
(Brandt et~al. 1997). 
%

In order to increase the statistics of hard X--ray data 
for NLS1 galaxies we have undertaken a program of 
BeppoSAX observations of a sizeable sample 
of NLS1 (about 12 over 3 years). We have selected, for the first 
year of observations, among the brightest and
most variable NLS1 previously observed by ROSAT and/or ASCA.
The spectral capabilities of the detectors onboard BeppoSAX
and especially the relatively large effective area at high energy ($>$ 2 keV) 
will allow a better study of the high energy properties of NLS1.

In this paper we present the first simultaneous 0.1--10 keV spectrum 
of the narrow-line quasar Tonantzintla~S~180 (hereafter Ton~S~180)
obtained with the BeppoSAX satellite,
together with a high-quality optical spectrum obtained 
at La Silla two months before the X--ray observation.

Ton S 180 is an optically bright (V = 14.4) galaxy with an optical spectrum 
typical of NLS1 at $z$ = 0.062. 
Multicolor UBVRI photometry revealed small amplitude
variations on the order of 0.1 mag over a timescale of a few months
(Winkler et al. 1992).
It is also a relatively bright extreme ultraviolet (EUV) source detected by
the EUVE satellite with a flux density  $\nu F_{\nu} \sim$ 3.9 $\cdot 10^{-11}$
erg cm$^{-2}$ s$^{-1}$ at 0.14 keV assuming a power law
spectrum of photon index $\Gamma$ = 2.4 and Galactic absorption
(Vennes et al. 1995). 
Large amplitude EUV variability has been also detected with a doubling time
of about 12 hours (Hwang \& Bowyer 1997).

Several PSPC observations, both from the ROSAT All-Sky Survey and 
the ROSAT pointed program, have been discussed by Fink et al. (1996).
The soft X--ray flux shows variability with an amplitude up to a
factor of $\approx 2$ on time scales ranging from hours to days and years.
The intensity variations are not clearly associated with significant
spectral variations. The $\sim$ 0.1--2.4 keV spectrum  
is steep, with photon index values ranging between 
$\sim$ 2.7 and 3.2.
When averaged over all the observations the mean PSPC spectrum 
is well represented  by a two component model consisting of a 
power law with $\Gamma$ = 3.10 $\pm$ 0.05 and a very steep low energy 
component, dominant below 0.3 keV, which can be represented 
either as a steeper power law or as blackbody emission at low temperature 
($\sim$ 16 $\pm$ 3 eV).

\section{The optical spectrum}

\subsection{Observations and reduction}

Optical spectra of Ton S 180 were obtained at the ESO 1.52m telescope on 1996
October 1 and~2, using the Boller \& Chivens spectrograph with a 127mm camera
and a Loral/Lesser thinned CCD with $2048\times2048$ pixels.  The pixel size is
$15\mu\hbox{m}$, and the projected scale on the detector is
0.82~$''$~pixel$^{-1}$.  The grating used had 600~grooves~mm$^{-1}$.  The
spectra were obtained through a 2~$''$ wide slit at a resolution of 4.6~\AA.
Both nights were photometric. Four integrations of 1800 seconds each were 
obtained.

Standard techniques were used to reduce the spectra, using the NOAO IRAF
package.  In particular, the spectrophotometric calibration of the spectra was
perfected by comparison with short
exposures of Ton~S~180 obtained during the same nights through an 8~arcsec slit.
This allowed us to correct the narrow-slit spectra for light losses and
differential refraction.  The four individual spectra were finally averaged and
yielded the final spectrum shown in Fig.~1.  Because the calibration curves
of the two nights and the individual spectra coincide within $<$5\%, we
consider the spectrophotometric quality of the final spectrum to be of this 
order.

\begin{figure*}
\epsfig{file=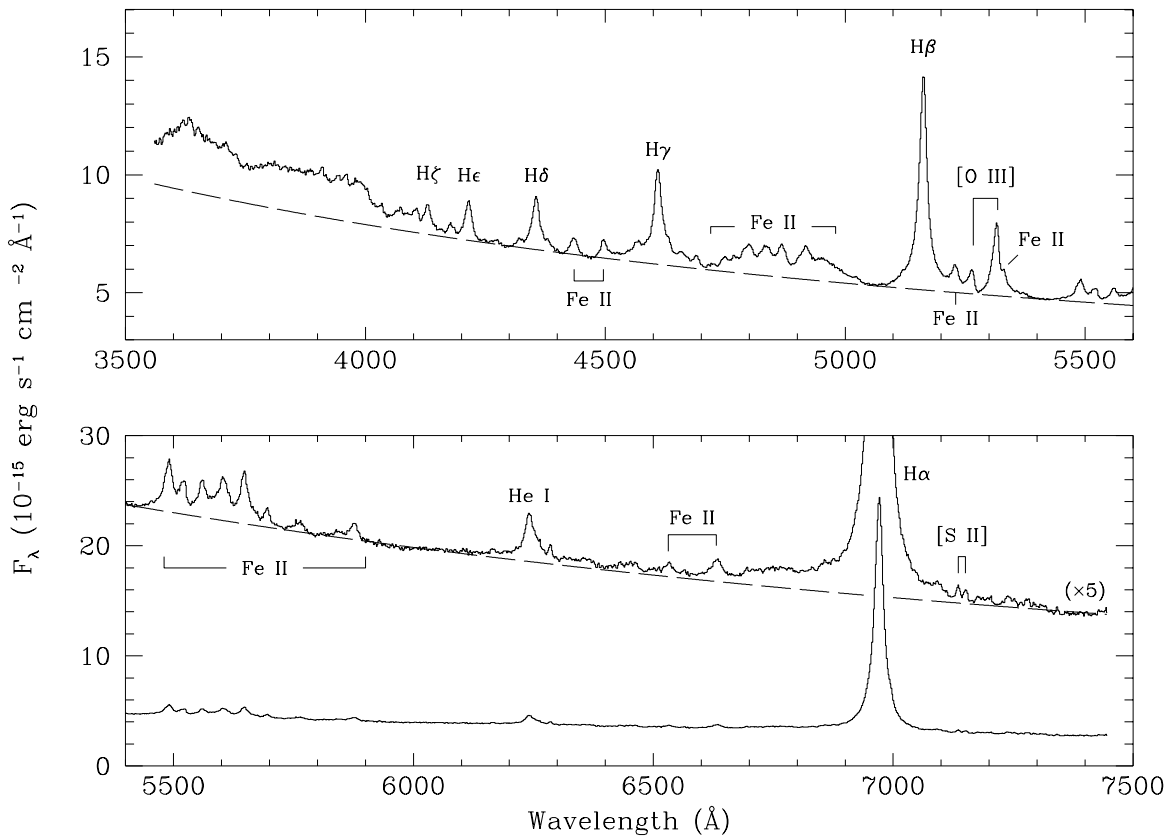, height=12.cm, angle=-90, silent=}
\caption{The optical spectrum of Ton S 180 obtained at ESO in 1996 October. 
The spectrum has been split in two segments for viewing purposes. 
The bottom panel also shows the red segment scaled up by a factor 5, 
in order to show the weaker lines. 
The main emission lines are indicated on the figure. The dashed line is the 
fitted power-law continuum (see text). The wavelength scale is in the rest 
system of the observer (z=0.062)}
\end{figure*}

\subsection{Properties of the optical spectrum}

The spectrum of Ton~S~180 presents characteristics normally associated with 
NLS1, such as strong Fe~{\sc ii} multiplets and a weak narrow-line spectrum. 
The only clearly visible forbidden 
lines are [O~{\sc iii}]$\lambda5007$ and $\lambda4959$, 
while [S~{\sc ii}]$\lambda6717$ and $\lambda6731$ are barely detected (Fig.~1).
There is no detection of other forbidden lines, such as those of [N~{\sc ii}], 
[O~{\sc i}], 
and [Fe~{\sc vii}], which are usually clearly visible in Seyfert spectra. 
The broad line spectrum of Fe~{\sc ii}, on the other hand, 
with the individual lines of multiplets 
27, 28, 37, 38, 42, 43, 46, 48, 49, 55, and 74 
is clearly visible (compare with Phillips 1976).

The continuum redwards of $\sim4300$~\AA\ is well represented by a power-law
function with slope $\alpha=-1.7$, where $F_\lambda\propto\lambda^\alpha$,
corresponding to $F_\nu\propto\nu^{-0.3}$.
The estimate of $\alpha$ was obtained by fitting a power-law to several dips in
between the emission lines. Figure 1 includes the fitted continuum. The strong 
excess emission at short wavelengths is part of the well-known small blue bump, 
which is centred at a rest wavelength of $\sim 3300$~\AA\ and is most likely the 
result of a blend between the Balmer continuum and several Fe~{\sc ii} 
multiplets.

The high signal-to-noise ratio of the spectrum allows to detect very broad, 
low-intensity wings under H$\alpha$, covering the $\sim6600$--7150~\AA\ range. 
Rodr\'\i{}guez-Pascual et al.\ (1997) have recently presented evidence for the 
presence of broad wings in the UV lines of a sample of NLS1 (which however did 
not include Ton S 180): there is therefore a possibility that similar components 
are present in the optical permitted lines (note however that 
Rodr\'\i{}guez-Pascual et al.\ searched for broad wings under H$\alpha$ in two 
objects of their sample without detecting them). While we do not exclude this 
possibility, we are also concerned that the feature, which is very 
blue-asymmetric, may be caused by the presence of Fe~{\sc ii} multiplets 
bluewards of H$\alpha$. Some lines of multiplet 74, in particular, are clearly 
visible. In addition a local deviation of the continuum from the chosen 
power-law may contribute to creating a spurious wing. More detailed discussion 
and analysis of this feature are deferred to a future paper. 

\begin{table}
\caption{Fluxes and widths of the main optical emission features}

\begin{tabular}{lll}
Feature$^a$  &  Flux$^b$ & FWHM$^c$ \\
\hline  
Fe II $\lambda$4570 blend  &  2.2 & ... \\
H$\beta$  &  2.6 & 1120 \\
 $[O III] \lambda 5007$  & 0.6 & 860 \\
Fe II $\lambda$5250 blend &  1.1 & ... \\
He I $\lambda$ 5876 & 0.2  & 1100 \\
H$\alpha$     & 7.1 & 950 \\
\hline 
\end{tabular}

$^a$ All wavelengths are in rest frame \\
$^b$ Units of $10^{-13}$ erg cm$^{-2}$ s$^{-1}$ \\
$^c$ km s$^{-1}$ \\

\end{table}

Table 1 lists the integrated fluxes of the main lines, derived by 
integrating the flux above a continuum fitted to the local minima close to the 
measured emission features. The strength of the Fe~{\sc ii} blends relative to 
H$\beta$, while not as high as that of I~Zw~1, the prototypical NLS1, is 
nevertheless much higher than that typical for normal Seyfert~1 nuclei (Joly 
1988).

The fluxes of H$\beta$ and [O~{\sc iii}]$\lambda5007$ were measured after 
subtraction
of the strongest blending lines with suitable templates.  
Fe~{\sc ii}~$\lambda4924$ and Fe~{\sc ii}~$\lambda5018$ 
were subtracted by using the profile of H$\alpha$
scaled by a factor 0.04, under the reasonable assumption that the lines in
multiplet 42 are of equal intensity.  
The scaling factor was chosen visually by trial and error, on the basis of
the smoothest residual spectrum.
To subtract [O~{\sc iii}]$\lambda4959$ we used
the deblended profile of [O~{\sc iii}]$\lambda5007$ scaled by one third.  
The ratio of H$\alpha$ and H$\beta$ (2.7) is one of the lowest observed in a
type~1 AGN. According to photoionization models based on typical
properties inferred for the broad-line region (BLR), 
and as low densities are excluded because of the relative narrowness of the
[O~{\sc iii}] lines, this may imply that 
densities in the BLR of Ton~S~180 are higher than $\sim 10^{11}$ cm$^{-3}$
(e.g.  Rees et~al. 1989, Zheng \& Puetter 1990).

\section{BeppoSAX Data reduction}

The X-ray satellite BeppoSAX (Boella et al. 1997a), a program of the 
Italian Space Agency (ASI)
with participation of the Netherlands Agency for Aereospace Programs (NIVR),
includes 
four co--aligned Narrow Field Instruments (NFI): a Low Energy
Concentrator Spectrometer (LECS; Parmar et al. 1997), three 
Medium Energy Concentrator Spectrometers
(MECS; Boella et al. 1997b), a High Pressure Gas Scintillation Proportional 
Counter (HPGSPC), and a Phoswich Detector System (PDS).
The LECS and MECS have 
imaging capability and cover the 0.1--10 keV and 1.3--10 keV energy ranges
respectively; in the overlapping band the total effective area of the
MECS (which is $\sim$150 cm$^2$ at 6 keV)
is about three times that of the LECS.
The energy resolution is  $\sim$8\%, and the angular resolution is 
$\sim$1.2$^{\prime}$ (Half Power radius) at 6~keV for both instruments.
 
BeppoSAX observed the source with the NFI on 1996 December 3  
for about 25 ks effective time with MECS and 12 ks with LECS. 
The shorter LECS exposure is due to the switching off of the 
instrument over the illuminated Earth. The HPGSPC and the PDS were
not sensitive for this rather faint source, and thus data are presented 
only for LECS and MECS. 

LECS spectra have been extracted from a 9$^{\prime}$ radius region around 
the centroid of the source image, which allows the collection of 
about $\sim$ 95 \% of the photons in the carbon band (E $<$ 0.3 keV).
For the MECS the adopted extraction radius was 4$^{\prime}$ . 
The source radial profiles in both the LECS and MECS are consistent
with those expected from the PSF without any evidence of
extended X--ray emission.
The spectra from the three units have been equalized to the
MECS1 energy--PI relationship and added together. Since the MECS 
background is very low and stable\footnote{For information on the 
background and on data analysis in general see
{\sl http://www.sdc.asi.it/software/cookbook} }, 
we used
all data acquired with an angle, with respect to the Earth's limb,
higher than 5$^{\circ}$. 

We are interested in the high energy spectrum of this source and
since it is rather faint, and possibly very steep, background
subtraction plays a crucial role.  Background counts were accumulated
from regions surrounding the sources and compared with counts
accumulated from the same regions from blanksky observations. The
`local' and `blank sky' background counts were always within 10 \% of each
other for the three MECS detectors and for the LECS.  
We used in our spectral analysis the `blank sky' background 
extracted from the `blank sky' event files 
in the same regions as the source.

The background subtracted count rates are 0.127 $\pm$ 0.004 cts~s$^{-1}$
for LECS in the 0.1--6 keV band and 0.096 $\pm$ 0.002  cts~s$^{-1}$ for 
the 3 MECS in the 1.5--10 keV energy range. 

\section{Timing analysis}

Figure 2 shows the background subtracted light curves over the 0.1--3~keV 
(LECS) and 3--10.5~keV (MECS) energy ranges. The time shown in the X axis
is computed from the beginning of the observation. The source data have been
extracted from a circular region with a 4 arcmin radius for both instruments.
The background, derived from an outer region at about 10 arcmin from the
center in the same observation, has been corrected for an average vignetting
correction estimated from the analysis of a deep observation of a blank sky
area. The background contribution to the total counts in the 4 arcmin region
is $\sim 5\%$ in the LECS and $\sim 13\%$ in the MECS data.

The data points shown in Fig.~2 correspond to periods of continuous 
observation. The length of the observation in each orbit is indicated by
the horizontal lines, while the vertical lines show the 1$\sigma$ statistical
uncertainty. Both light curves show significant variability, of the order of
a factor of two, during the length of the entire observation. The hypothesis
of a constant count rate can be rejected with a high level of confidence 
in both the 0.1--3.0~keV (p $<$ 1 $\cdot 10^{-5}$) and 3.0-10.5~keV
(p $\sim$ 8 $\cdot 10^{-4}$) energy ranges. 
No significant variability is instead detected within the data from 
single orbits (i.e. on timescales of the order of 1,000 -- 3,000 s).

Visual inspection of the two light curves suggests the possibility
that, during the increase of the source flux (at t $\sim$ 10,000 s),
the hard curve lags the soft one. In fact, the peak in the MECS data is seen at
t $\sim$ 17,000 s, while the correponding peak in the LECS data is seen at
t $\sim$ 12,000 s. However, the significance of this effect is only
marginal and data with better statistics and/or with a more continuous time
coverage would be needed to reliably conclude that this time lag is real.
For example, the two solid lines in Fig.~2 show two light curves which, having
been obtained by keeping fixed the ratio between
LECS and MECS counts (i.e. with no spectral variation), are however consistent
at better than 2$\sigma$ with both the LECS and MECS data. The same conclusion
is reached by analyzing the hardness ratio, defined as the ratio between
the MECS and LECS counts, as a function of time. Therefore, since the data
are consistent with a spectrum which changes only in normalization,
in the next section we will derive the shape parameters for the spectrum using
the data averaged over the entire observation.

\begin{figure}
\epsfig{file=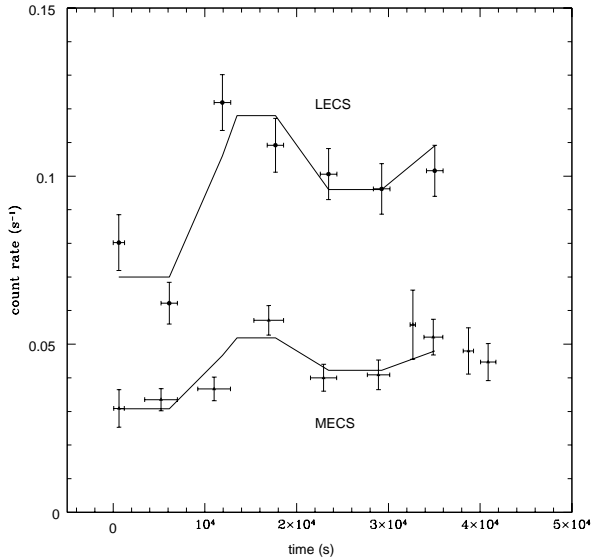, height=8.cm, angle=0, silent=}
\caption{Background subtracted LECS (0.1--3~keV) and MECS (3--10.5~keV)
light curves, obtained from a circular region with a 4 arcmin radius for both
instruments. The meaning of the two solid lines is described in the text}

\end{figure}

\section{Spectral analysis}

\subsection{The 0.1--10 keV spectrum}

For the spectral analysis the source counts have been grouped with a minimum
of 20 counts per energy bin in the $\sim$ 0.1--6 keV (LECS) and 
$\sim$ 1.5--10 keV (MECS) energy ranges. The LECS  and MECS datasets 
have been fitted simultaneously with the relative normalizations free to vary,
in order to take into account the different exposure times and the
relative calibration of the two instruments (Cusumano et al. in preparation)
The spectral fits have been performed with the 
XSPEC 9.0 fitting package, using the response matrices released on Jan
1997. In the following, all quoted errors correspond to the 
90\% confidence level.
A summary of the results is reported in Table~2.

The first line of Table 2 shows the parameters obtained for the
best fit model. This model, computed by keeping the $N_H$ value fixed
at the Galactic value measured in the direction of Ton S 180 ( (1.5 $\pm$ 0.5)
$\times~10^{20}$ cm$^{-2}$; Dickey \& Lockman 1990), is characterized
by two power laws, with best fit slopes $\Gamma_s$ = 2.68 and $\Gamma_h$ =
2.29, plus a narrow ($\sigma \equiv$ 0.01 keV) gaussian line centered at 
7.11 $\pm$ 0.16 keV with
an equivalent width of 507 $\pm$ 247 eV. The resulting fit ($\chi^2$ =
135 with 135 degrees of freedom) is shown, superimposed on the data, in 
the upper panel of Fig.~3. The lower panel, showing the ratio between
data and model, clearly indicates that this model is a good representation
of the data over the entire energy range. Similar results are obtained by
leaving the $N_H$ value free to vary, with a best fit value of $N_H$ = 
(1.15 $\pm$ 0.45) $\times~10^{20}$ cm$^{-2}$, consistent with the
galactic value. Substituting the power law at low energy
with a blackbody emission produces a significantly worse fit quality
($\Delta\chi^2$=8.1 with the same number of degrees of freedom; see second 
line in Table 2).

\begin{figure}
\epsfig{file=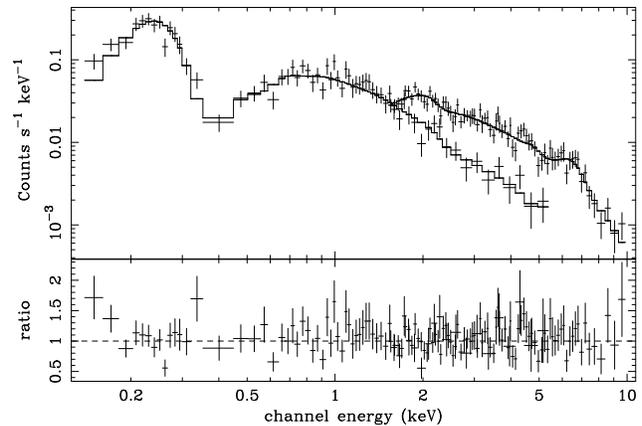, height=7.cm, angle=-90, silent=}
\caption{The LECS + MECS spectrum of Ton~S~180. 
The energy scale is in the rest system of the observer.
See the text for the 
spectral model that has been fitted to the data}
\end{figure}

The line energy is consistent at the 90 \% level with a hydrogen--like 
iron line at 6.97 keV. Leaving the width $\sigma$ of the line free to vary the
improvement in the fit quality is only marginal ($\Delta \chi^2 \simeq 2.5$).
The formal width is $\sim$ 0.5 keV and the gaussian centroid is $\sim$ 6.9 keV.
The best fit line energy is therefore suggestive of a highly ionized
accretion disk. A characteristic strong feature from such a disk 
is expected to be a smeared iron absorption edge 
(e.g. Ross et~al. 1996), but this feature 
is not seen in the present data.
Alternatively, the $\sim$ 7.1~keV line could be  
the blue horn of a relativistic disk line from an edge--on disk,
rather than a ionized one. 
To test this possibility we have fitted the diskline model in XSPEC 
to the observed spectrum. 
Unfortunately, even though an acceptable description of the data can be
obtained with a disk line, the quality of the present data does not allow us
to put strong constraints on the inclination angle.

We have also tried to model the soft X--ray spectrum below $\sim$ 2 keV
with more complex models fixing the absorption at the Galactic value.
Both an ultrasoft blackbody plus a power law and a double blackbody
(which would be a first--order approximation of the accretion disk emission)
provide acceptable descriptions of the soft X--ray spectrum.
However, they are not statistically required by the present data.

On the basis of the best fit model, the unabsorbed 0.1--2.0~keV flux is
$\sim 3 \cdot 10^{-11}$ erg cm$^{-2}$ s$^{-1}$, corresponding to a soft
X--ray luminosity of $\sim  5.3 \cdot 10^{44}$ erg s$^{-1}$ 
($H_0$ = 50 km s$^{-1}$ Mpc$^{-1}$; $q_0$ = 0).
This places Ton~S~180 among the X--ray luminous quasars.
The soft X--ray flux is almost an order of magnitude greater 
than the observed 2--10~keV flux of $\sim 4.2 \times 10^{-12}$ 
erg cm$^{-2}$ s$^{-1}$, which corresponds to a 2--10 keV luminosity
of $\sim 7.4 \cdot 10^{43}$ erg s$^{-1}$.

Simpler models (i.e. with a smaller number of free parameters) do not adequately
fit the data over the entire range of energy. This is shown in the next
two lines of Table 2, which give the best fit parameters and the
$\chi^2$ values of the fits for a model with two power laws and no emission
line (third line in Table 2) and a single power law with emission
line (fourth line in Table 2). In both cases, the quality of the fit, as
judged by the increase in the $\chi^2$ value, is significantly worse than
that obtained for our best fit model. In the first case a clear excess in
the residual is present around 6--7 keV (Fig.~4); in the second case 
systematic residuals both at low and high energies are left, moreover the
extremely high value found for the iron line 
equivalent width also suggest a rather poor approximation 
of the continuum at energies $>$ 5--6 keV.

\begin{figure}
\epsfig{file=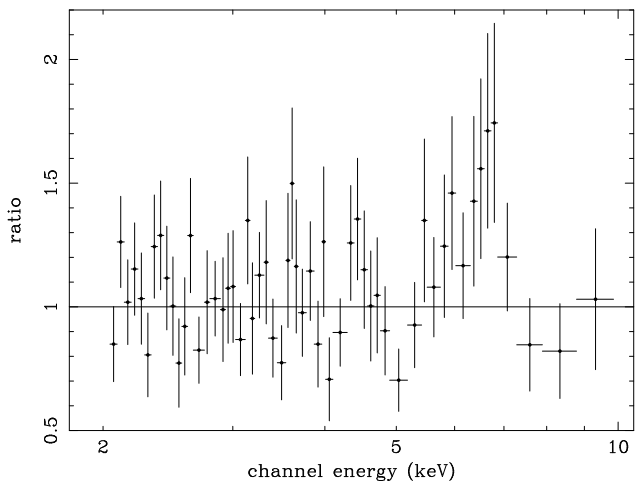, height=7.cm, angle=-90, silent=}
\caption{Residuals from a single power law fit ($\Gamma = 2.21\pm0.12$) 
in the 2--10~keV band. The energy scale is in the rest system of the observer}
\end{figure}

\subsection{Comparison with ROSAT results}

The direct comparison of the spectral analysis results with those obtained
by Fink et al. (1996) is not straightforward given the source variability 
and the different assumptions about the continuum model between ROSAT and 
BeppoSAX. The range of PSPC spectral slopes is in good agreement
with the power-law fits of the soft component reported here (Table~2).
This suggests that the slope of the soft X--ray spectrum remains steep
on timescales of several years, while the soft flux undergoes significant 
and rapid variations.
The unabsorbed 0.1--2.4 keV flux of  2.6 $\cdot 10^{-10}$ erg
cm$^{-2}$ s$^{-1}$, derived by Fink et al. (1996)
on the basis of a two component model consisting of an ultra--soft blackbody
plus a steep power law, is almost an order of magnitude greater than
the present measurement. While this could be due, at least in part, to the 
different
adopted spectral parameters and in particular to the larger value
of the column density derived by Fink et al. (1996), 
it is likely that Ton S 180 was, 
on average, brighter during the ROSAT observation.
The derived flux density at 0.14 keV of 
1.5--1.8 $\cdot 10^{-11}$ erg cm$^{-2}$ s$^{-1}$, depending on the assumed
spectral model, is more than a factor 2 lower than
that obtained by EUVE (Vennes et al. 1995), for a similar assumption on 
the spectral parameters, confirming large amplitude flux variability.

\begin{table*}
\caption{LECS+MECS joint fits in the 0.1--10 keV energy range}

\vglue0.3truecm
{\hfill\begin{tabular}{l l l l l l l}
 $F_{1~keV}$ & $\Gamma_s$ & $E_{break}$  & $\Gamma_h$ & $E_{line}$ & EW & 
$\chi^2$/d.o.f. \\
  $\mu$Jy  & $kT$ (eV) &     keV        &   &  keV & eV &            \\
&&\\
\hline  
 2.16 & 2.68$\pm$0.07  & 2.5$^{+0.5}_{-0.9}$ & 2.29$^{+0.12}_{-0.19}$ &
 7.11$\pm$0.16 & 507$\pm$247 & 135.0/135 \\
&&\\
 1.95  & 23$^{+18}_{-12}$  & ... & 2.48$\pm$0.08 & 7.11$\pm$0.14 & 
750$^{+240}_{-310}$ & 143.1/135 \\
&&\\
 2.16 & 2.68$\pm$0.07  & 2.5$\pm$0.7 & 2.16$\pm$0.16 & ... & ... & 
146.0/137 \\
&&\\
 2.20 & 2.59$\pm$0.06  & ... & ... & 7.11$\pm$0.14 & 887$\pm$310 & 
157.6/137 \\
&&\\
\hline  
 
\end{tabular}\hfill}

The absorption column density has been fixed at the Galactic value in all 
the fits.\\

\end{table*}

\section{Discussion} 

The X--ray spectral properties of Ton~S~180 are remarkable 
when compared with those of classical Seyfert~1s and quasars.
First of all, the best fit centroid energy of the iron line 
and its large equivalent width suggests emission
from optically thick ionized matter (e.g. an accretion disk).
The properties of iron lines from ionized disks
have been studied in detail by Matt et~al. (1993, 1996) and \.Zycki
\&  Czerny (1994). The total equivalent width of the line, if the iron
is predominantly in the He--like and H--like states, can be much higher (up to 
more than seven times) than when the iron is neutral
(see e.g. Fig.~2 of Matt et~al. 1996). This is due 
to the increased fluorescent yield and the
decreased photoelectric absorption opacity. If only 
the non-scattered component (which is the 
only one easily detectable) is considered, the EW can be as high as 3 times 
that of neutral iron. The observed EW, i.e. $\sim$500$\pm$250 eV, is then
consistent with the theoretical calculations. 
%
The present spectra do not allow us to investigate
the line profile in detail. Therefore, in principle the feature at 
$\sim$~7~keV could also be the blue horn of a relativistic line from an 
approximately edge--on accretion disk. It has been proposed 
that relativistic effects from an approximately edge--on 
accretion disk could provide the boosting needed to explain 
the giant-amplitude variability observed in the NLS1 IRAS~13224--3809 
(Boller et~al. 1997). An edge--on disk near a Kerr black hole can 
also lead to the strong soft X-ray excess often seen in NLS1 
(see Cunningham 1975 and the 
discussion in Sect.~5.1.7 of Boller et~al. 1997). 
A difficulty with this explanation is the observed line
equivalent width, which is an order
of magnitude larger than expected from an edge--on disk
(Matt et al. 1992).

A two-component model provides an adequate description of the
continuum X--ray spectrum of Ton~S~180, in analogy with several broad 
lined Seyfert~1s and quasars. However, the slopes
of the two components are rather different from those in broad-lined 
Seyfert~1s (Comastri et al. 1992; Walter \& Fink 1993).
The soft component is much stronger than in ``normal" 
Seyfert~1s, while the 2--10~keV slope is steeper 
than the average value observed in Seyfert~1 galaxies (Nandra \& Pounds 1994).  
The increasing evidence for similar behaviour in other
NLS1 (Brandt et~al. 1997) raises
the problem of the origin of such a different spectrum and its relation
to the optical properties.
The large relative intensity of the soft component, which in Ton S 180
is about a factor 7--8 times greater than the 2--10 keV one, 
and the steep 2--10 keV slope have important 
consequences for various models. 
Even if the observation of a strong iron line suggests that reprocessing
is occuring at some level, it appears likely that the strong soft component 
cannot be due only to disk reprocessing. One would require
either highly anisotropic emission or a high energy spectrum 
extending up to several tens of MeV without any cut--off, at variance
with recent high energy observations of Seyferts 
(e.g. Gondek et al. 1996).
The hint of a possible lag of the hard photons with respect to the soft ones
(see Sect.~4 and Fig.~2) also argues against disk reprocessing
as in this case the soft component is expected to lag the hard one.

A soft X--ray component much stronger than expected 
from a reprocessing origin alone could lead to 
a strong Compton cooling of the hot corona electrons and so 
to a steep hard tail 
if a significant fraction of the gravitational 
energy is dissipated in the disk phase. 
This hypothesis is also supported by the similarities
between the X-ray spectra of RE~J~1034+393
(Pounds et~al. 1995) and Ton S 180, with those of Galactic 
black hole candidates (GBHC) in their high states.
The high states of GBHC are thought to be triggered by increases in the
accretion rate resulting in strong thermal emission 
from a disk accreting at the Eddington 
limit (e.g. Tanaka 1990; Ebisawa 1991). 
It is worth noticing that if the black hole is accreting near the
Eddington limit, then the disk surface is expected to be strongly ionized, 
which fits nicely with the present detection of an ionized line. 
Ionized iron line emission has been recently discovered by ASCA  
among high luminosity quasars (Nandra et al. 1996,1997),
indicating that these sources are radiating at a substantial fraction of the
Eddington limit. The present results corroborate the hypothesis of 
a high L/L$_{Edd}$ ratio in Ton~S~180.
Note that this is also in agreement with the 
proposed explanation of the narrow-line Seyfert~1s phenomenon in terms 
of a high L/L$_{Edd}$ ratio (BBF96; Czerny et al. 1996; Laor et al. 1997):  
the optical line width will be inversely proportional 
to L/L$_{Edd}$ if the broad line region is virialized.

If the soft X--ray component in Ton~S~180 is not 
the high energy tail of a hot disk accreting at the Eddington limit, 
but is produced by Comptonization of a 
lower temperature disk (e.g. Haardt \& Maraschi 1993), 
which would be consistent with the good fit obtained with a power law 
in the 0.1--2.0~keV range (Table~2), then the origin of 
the steep 2--10~keV component would remain unexplained, 
unless a more complicated two temperature comptonizing corona 
is assumed.

The optical and X--ray spectral energy density distribution 
is shown in Fig.~5. A simple extrapolation of the data suggest that 
the energy density
peaks somewhere in the extreme ultraviolet, with a behaviour 
similar to that of other 
quasars (e.g. Zheng et al. 1997). Also the broad band  
spectral index $\alpha_{ox}$ = 1.53 
between 3000~\AA\ and 2 keV , is close to the average value
of 1.56$\pm$0.26 of the radio--quiet quasars in the  Laor et al. (1997)
sample.  

\begin{figure}
\epsfig{file=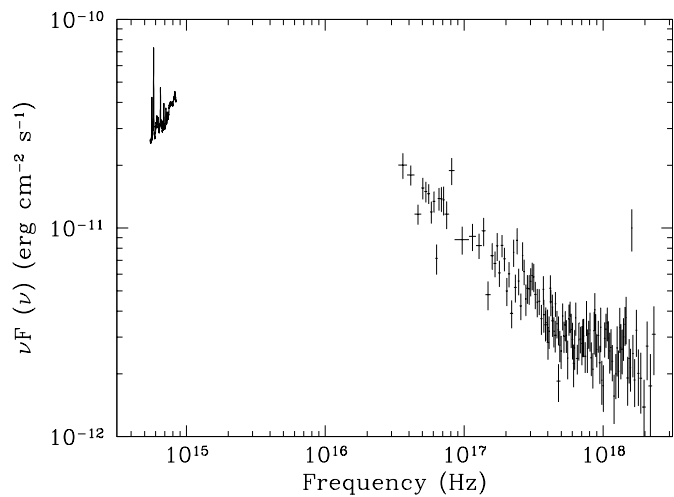, height=7.cm, angle=-90, silent=}
\caption{Spectral energy distribution in the optical and X--ray bands}
\end{figure}
 
Given the large-amplitude
variability of Ton~S~180, the lack of simultaneous 
multifrequency observations, and the difficulties that accretion disk 
models have explaining both the Optical--UV and the soft X--ray
emission of AGNs (e.g. Ulrich \& Molendi 
1995; Siemiginowska et al. 1995; Laor et al. 1997), 
no attempt has been made to fit the observed spectral energy distribution 
of Fig.~5 with a disk model.

The link between X--rays and optical properties needs also further
examinations. While Ton S 180, similarly to most of the NLS1, 
displays strong Fe {\sc ii} emission (Fig.~1) 
and relatively rapid X--ray variability, the other well studied NLS1 
RE~J~1034+393, with a similar X--ray spectrum and a ionized iron line
(Pounds et~al. 1995), does not show 
significant evidences of X--ray variability and displays weaker
Fe {\sc ii} lines. 
The steep 2--10 keV slope coupled with the strong optical Fe {\sc ii}
emission in Ton~S~180 provide evidence against models of optical  
Fe {\sc ii} line
formation that require flat X--ray spectra and strong hard X--ray 
emission (e.g. Collin--Souffrin et~al. 1988).

Further BeppoSAX observations and coordinated optical campaigns 
are clearly needed to clarify some of these issues. 
A more detailed discussion of the X--ray spectral properties of NLS1 
is postponed until the observational program is completed.

\section{Summary}

We have analysed BeppoSAX data and a high-quality optical spectrum
of the bright narrow-line Seyfert~1 galaxy Ton~S~180.
Our main results are the following:

(1) The optical spectrum clearly confirms
the NLS1 nature of Ton~S~180 and reveals strong Fe~{\sc ii} 
emission. Moreover, the H$\alpha$ to H$\beta$ ratio, which is one of the lowest
measured among AGNs, may imply higher densities than usual in the line 
emitting region.

(2) Large-amplitude variability of about a factor 2 
in both the soft (0.1--3.0~keV) and medium
(3.0--10.5~keV) energy ranges has been detected. 

(3) The 0.1--10~keV spectrum requires at least two components: (1) a steep
$\Gamma \simeq$ 2.7 strong soft component below 2~keV which contains a 
large fraction of the
overall energy output, and (2) a weak hard tail with a 2--10~keV slope
$\Gamma \simeq$ 2.3 
that is significantly steeper than the average value found in 
``normal'' Seyfert~1s and quasars.

(4) There is evidence for iron line emission at $\sim$ 7~keV. When fitted with
a simple Gaussian, the best fit centroid indicates a high ionization state.
Together with the strong soft excess, the steep high energy 
slope, and the narrow optical lines, these observations suggest that 
the source is accreting near the Eddington limit and that most of the power 
is dissipated in the cold disk rather than in the hot corona.

(5) If, alternatively, the observed iron feature is the blue horn of a line 
produced in a relativistic accretion disk seen edge--on, then 
significant high amplitude, or even giant, variability is expected due to
relativistic boosting effects near the black hole.
Optical--UV  and X--ray monitoring as well as a broad band deep 
spectral observation of Ton S 180 appears to be very promising for a better 
understanding of the physics of  NLS1.

\begin{acknowledgements}
We thank all the people who, at all levels, have made possible the SAX mission.
This research has made use of SAXDAS linearized and cleaned event
files (rev0.1) produced at the BeppoSAX Science Data Center. 
We thank F. Haardt for useful discussions. AC and GZ 
acknowledge financial support from the Italian Space Agency under
contract ASI--95--RS--152.
\end{acknowledgements}

\end{document}